\documentclass[twocolumn,amssymb, nobibnotes, aps, prb]{revtex4-2}
\usepackage{amssymb}
\usepackage{appendix}
\usepackage{verbatim}
\usepackage{float}
\usepackage{color}
\usepackage{textcomp}
\usepackage{gensymb}
\usepackage{hyperref}
\usepackage{dsfont}
\hypersetup{
     colorlinks   = true,
     citecolor    = blue
}
\usepackage{siunitx}
\usepackage{graphicx}
\usepackage{dcolumn}
\usepackage{bm}
\usepackage{braket}
\usepackage{empheq}
\usepackage{nicefrac} 
\usepackage[dvipsnames]{xcolor}

\usepackage[normalem]{ulem}
\usepackage{balance}

\begin{document}

\title{Tunable large Berry dipole in strained twisted bilayer graphene}

\author{Pierre A. Pantale\'on$^1$}
\email{ppantaleon@uabc.edu.mx}
\author{Tony Low$^2$}
\author{Francisco Guinea$^{1,3}$}
\affiliation{$^1$Imdea Nanoscience, Faraday 9, 28047 Madrid, Spain}
\affiliation{$^2$Department of Electrical and Computer Engineering, University of Minnesota, Minneapolis, Minnesota 55455, USA}
\affiliation{$^3$ Donostia International Physics Center, Paseo Manuel de Lardiz\'abal 4, 20018 San Sebastián, Spain}

\begin{abstract}
Twisted bilayer graphene is highly sensitive to external perturbations. Strains, and the presence of the substrate, break the symmetries of the central bands. The resulting changes in the Berry curvature lead to valley currents, and to a non linear Hall effect. We show that these effects, described by a Berry dipole, can be very significant, such that the non linear effects surpass the linear response for moderate applied fields, $\sim 0.1 {\rm mV} / \mu \rm{m}$. The dependence of these effects on applied strain, coupling to the substrate, density of carriers, and temperature makes them highly tunable. 

\end{abstract}

\maketitle

\textit{Introduction.-} The observation of non-linear Hall effects in time-reversal-invariant transition-metal dichalcogenides~\citep{Ma2019,KangMak2019,huang2020giant,hu2020nonlinear} has sparked interest in different types of low dimensional materials~\citep{Araki2018, XiaoHua2020, FacioJeroen2018,Wang2019c,Du2019}. The non-linear Hall effect is a second order response to an in-plane electric field and does not need time-reversal symmetry breaking but requires an inversion symmetry breaking~\citep{TonyLowPaco2015,SodemannLiang2015}. The conventional Hall conductivity or linear Hall effect requires a broken time-reversal~\citep{XiaoBerry2010,NagaosaHall2010} because the Berry flux over the equilibrium distribution is zero if time reversal is present. However, the Berry curvature can emerge locally with counter-propagating charge carriers having different Berry curvatures. The non-linear Hall effect depends on higher order moments of the Berry curvature~\citep{SodemannLiang2015,NandySodemman2019}. Under an applied in-plane electric field, there is an imbalance between counterpropagating charge carriers, which in the presence of a dipolar distribution, or Berry dipole, generates a non-linear Hall current which scales quadratically with the electric field.  

\begin{figure}
\begin{centering}
\includegraphics[scale=0.40]{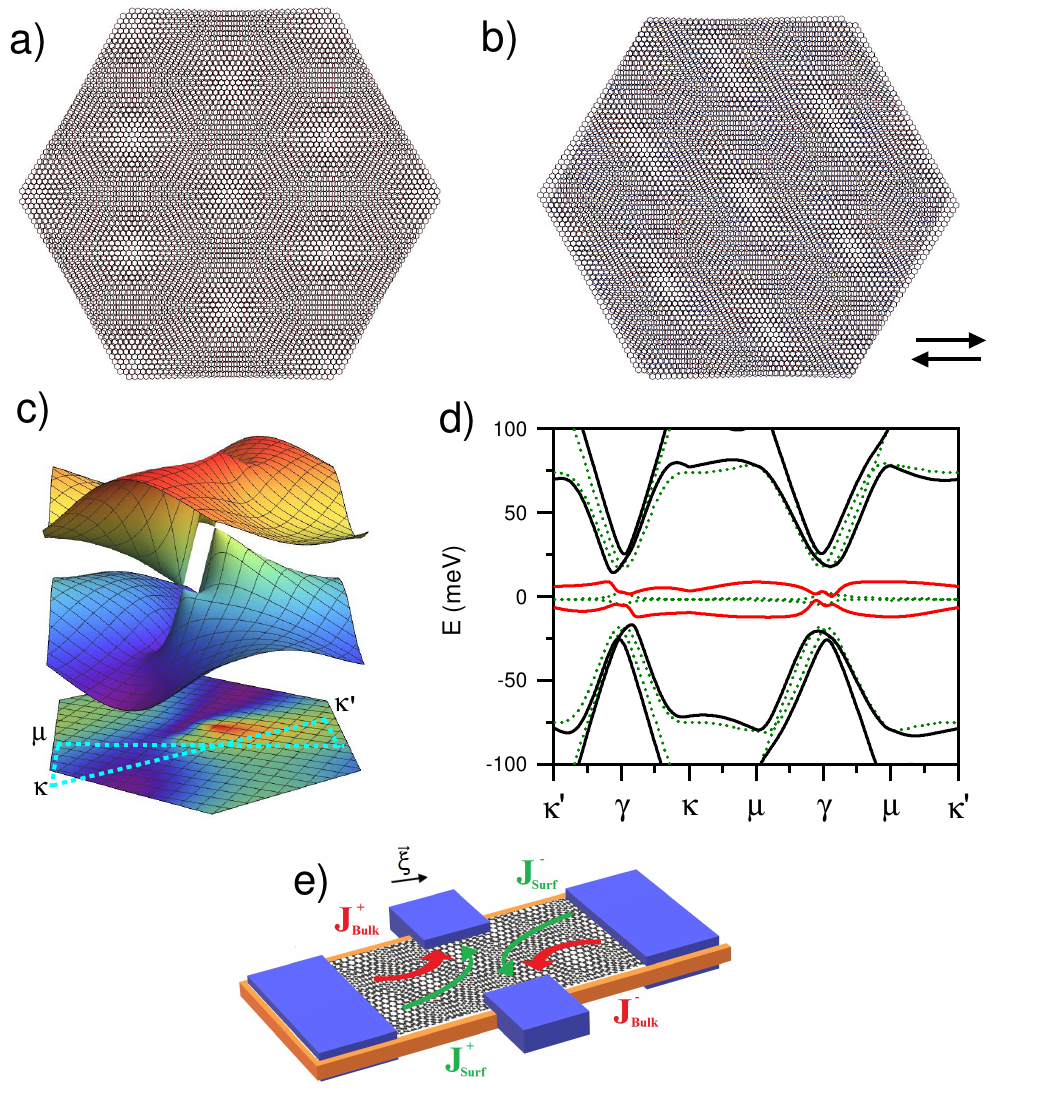}
\par\end{centering}
\caption{a) Sketch of the moir\'e superlattice for pristine TBG and b) strained TBG where elliptical moir\'e dots are induced, black arrows indicate the heterostrain direction \citep{SI}. In this situation, the c) band structure is distorted and the Dirac cones are not longer degenerate. The lower hexagon is the projection of the lower band over the moir\'e Brillouin zone. Color scheme is purple for the minimum and red for the maximum energy. Dotted path is used to calculate the d) band structure where we set a twist angle $\theta=1.05^{\circ}$ and uniaxial strain $\epsilon=0.3\%$, $\phi=0^{\circ}$. The staggered sublattice potential in each graphene layer is $\delta_{b}=\delta_{t}=0.01$ meV and the Chern numbers are $\mathcal{C}=\pm1$ for the lower and upper band, respectively (red lines) \citep{SI}. The band structure for the unstrained TBG is shown as green dotted lines. e) Illustration of the various transverse currents in a typical four terminal device. \label{fig: Figure1}}
\end{figure}

It has been recently shown that uniaxial strains enhance the Berry dipole~\citep{YouLow2018,ZhouZhangLaw2019,SonKimLee2019} in transition-metal dichalcogenides (TMD). In these materials, orbital valley magnetization~\citep{SonKimLee2019, ShiJustinMagnet2019}, giant magneto-optical effects~\citep{Liu2020} and non-linear Nernst effects~\citep{YuTony2019,ZengTewari2019}  can be induced as a response to an in-plane electric field due to the Berry dipole. We consider here the non linear Hall conductivity, and non local topological currents in twisted bilayer graphene (TBG). This system shows a wealth of unexpected properties, and its electronic bands have non trivial topological features~\cite{Cao2018,Cao2018_bis}.
In suspended or encapsulated twisted bilayer graphene with hexagonal Boron Nitride (hBN) the inversion symmetry is broken, resulting in narrow bands with a finite Berry curvature~\citep{Serlin2019,Song2015,Zhang2019a,bultinck_cm2019,Zhang2019b,CeaPantaGuinea2020}. However, recent experiments have mapped the strain fields in TBG~\citep{kazmierczak2020strain}, so it is natural to ask what is the magnitude of non-linear Hall effects in TBG. 

In this paper, we show that the large Berry curvature of the narrow bands in strained TBG leads to an strong and tunable non-linear Hall effect. The combined effect of the band topology and strain, generates a large Berry dipole. Importantly, the resulting non-linear current is  tunable by the strain parameters, electron fillings and temperature.

\begin{figure*}
\begin{centering}
\includegraphics[scale=0.053]{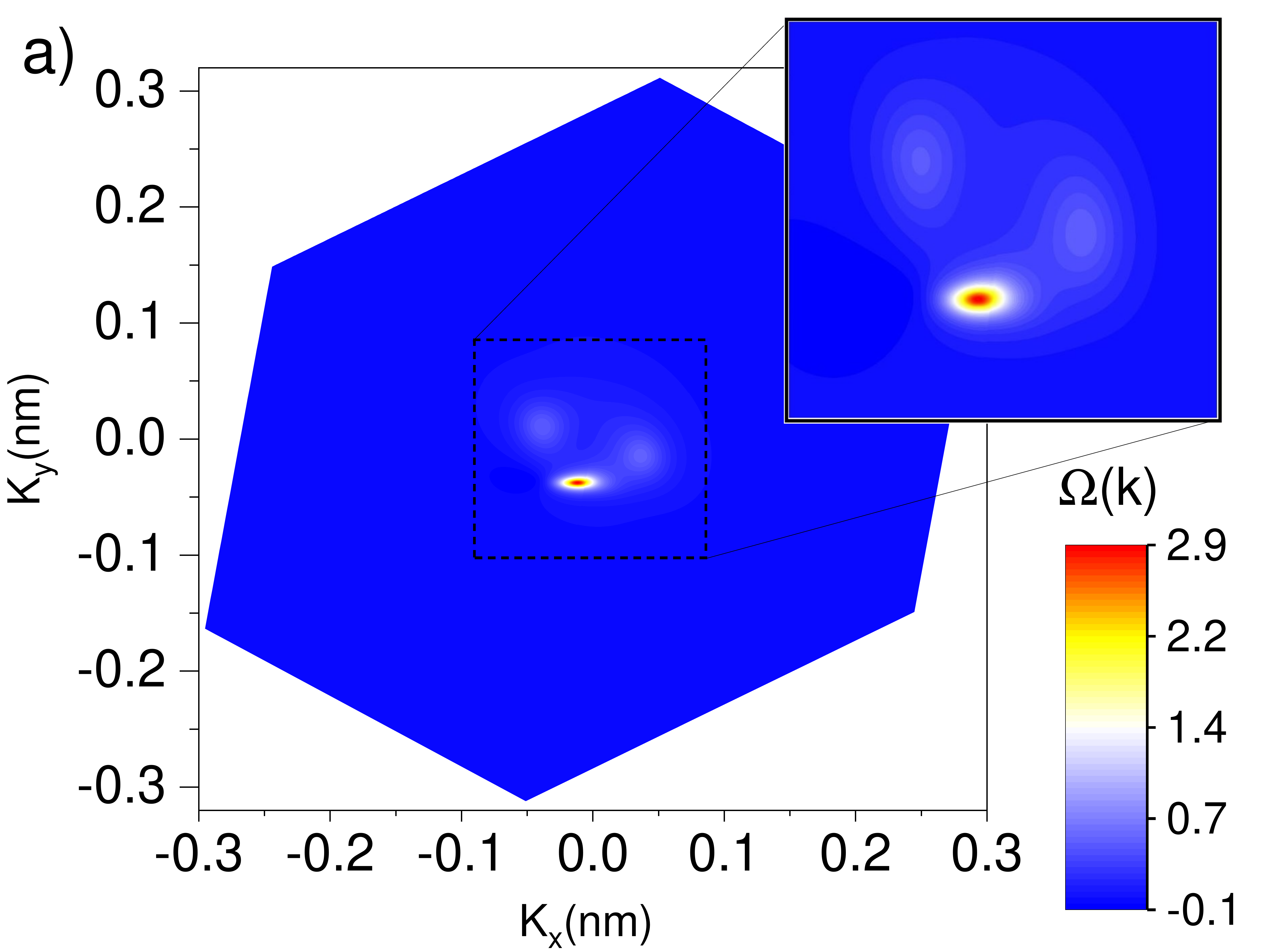} \includegraphics[scale=0.053]{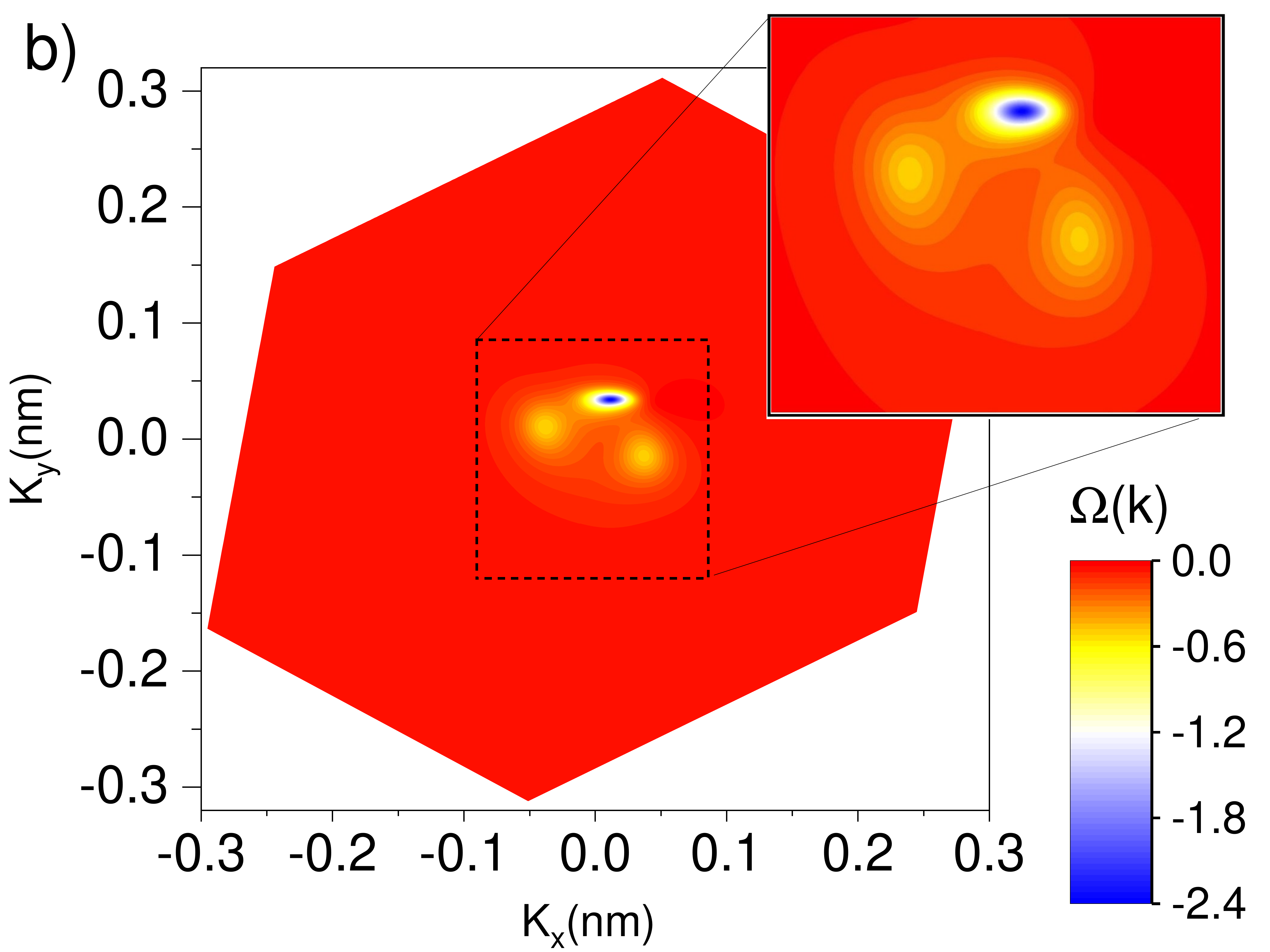} \includegraphics[scale=0.053]{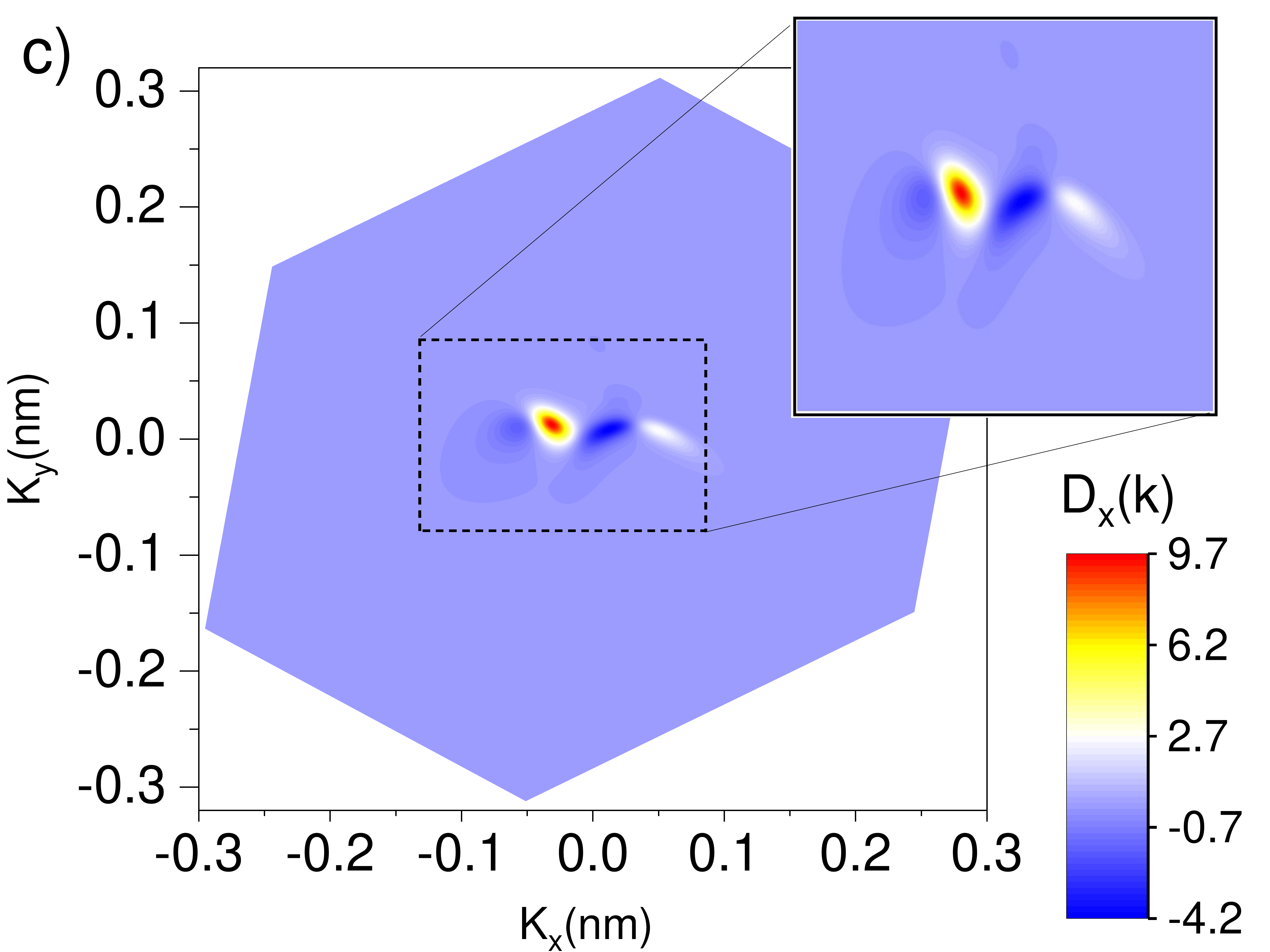} \includegraphics[scale=0.053]{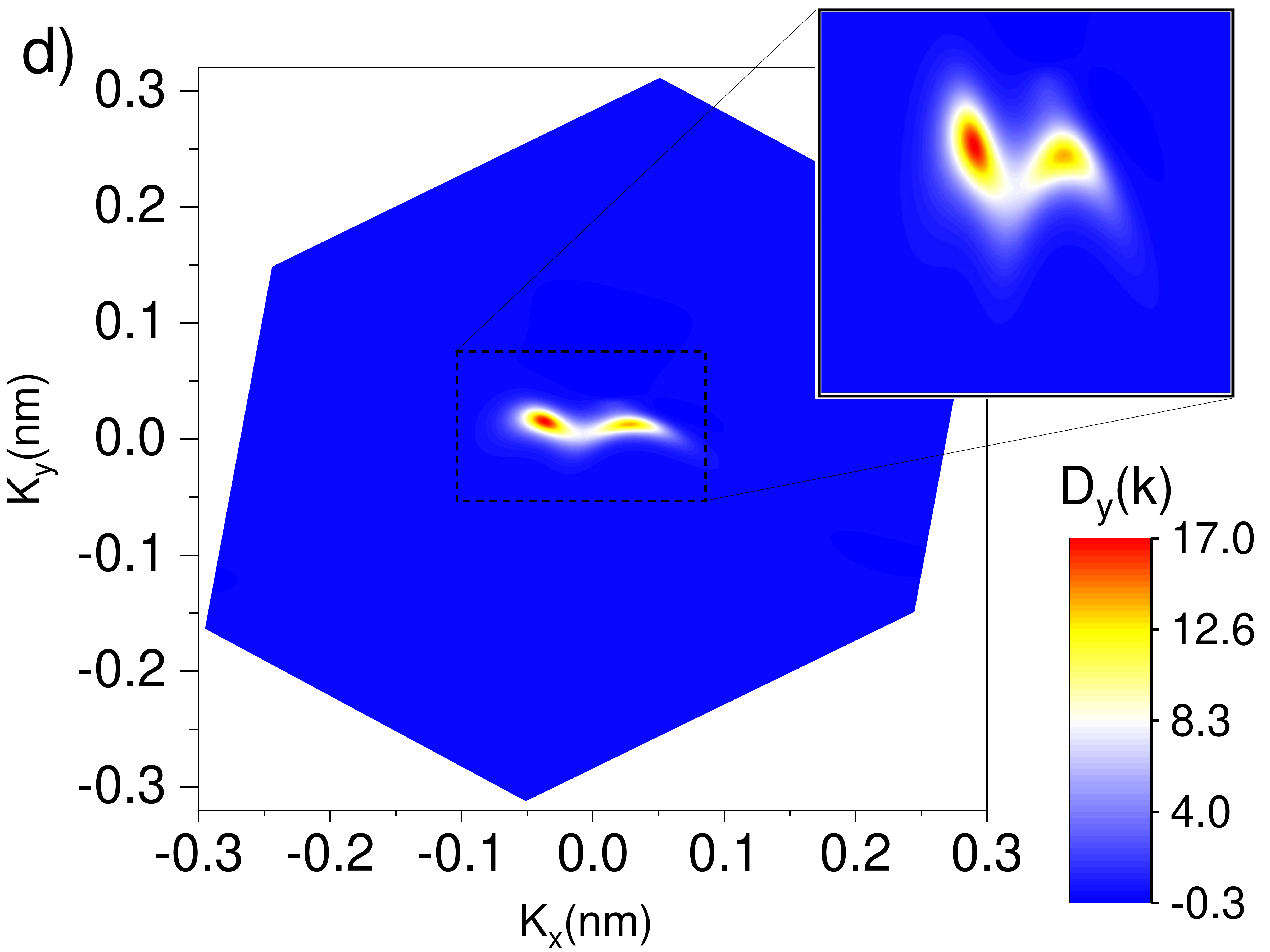}
\par\end{centering}
\caption{Berry curvature of the a) lower and b) upper narrow bands with Chern numbers $\mathcal{C}=\pm1$, respectively. For a chemical potential $\mu=5.6$ meV we plot the distribution of the c) $D_{x}$ and d) $D_{y}$ Berry dipole components in the mBZ. Parameters as in Fig. \ref{fig: Figure1} with $T=10 \text{ K}$ and $\delta_{b/t}=8.0 \text{ meV}$. Insets in each panel are the enlarged regions where the corresponding Berry curvature and distribution of the Berry dipole is concentrated. Berry curvature in a) and b) in units of $10^{-4}$ $\text{nm}^2$ and Berry dipole b)-c) in units of $10^{-4}$ $\text{nm}$.}  
\label{fig: Figure2a}
\end{figure*}

\textit{Topological currents.\textendash} Strains with opposite sign (heterostrain) in the two layers of twisted bilayer graphene of $0.1 - 0.5 \%$ have been measured in STM experiments~\cite{Jetal19,Choi2019,Xie2019,Kerelski2019,Quiao2018ht,Huder}. These strains can significantly distort the band structure, through two separate effects: i) the appearance of an effective gauge field of opposite signs in the two layers, and ii) a modification of the interlayer tunneling due to the changes in the stacking~\citep{BiFu2019} (see also\cite{SI}). Fig. [\ref{fig: Figure1}]a)-b) illustrate the moir\'e superlattice for pristine and strained TBG where elliptical moir\'e dots are induced. 

In pristine TBG both time reversal and inversion symmetry are preserved. In the presence of uniaxial heterostrain, the strain tensor breaks all the point group symmetries of the lattice except $\mathcal{C}_{2z}$. As a result, both narrow bands are still connected by two Dirac crossings. However, as shown in Fig.~[\ref{fig: Figure1}]c), these crossings are no longer at the corners of the moir\'e Brillouin zone (mBZ) and they are not at the same energy~\citep{BiFu2019}. In time-reversal invariant systems, the Berry curvature for the electronic Bloch states of the $n$th band, $\boldsymbol{\Omega}_{n}(\boldsymbol{k})=2 \mathrm{Im}\left\langle \partial_{k_{x}}\Psi_{nk}\right|\left.\partial_{k_{y}}\Psi_{nk}\right\rangle \hat{z}$ 
is odd in momentum space, that is $\Omega_{n}(\boldsymbol{k})=-\Omega_{n}(-\boldsymbol{k})$,
while a crystal lattice with inversion symmetry would require $\Omega_{n}(-\boldsymbol{k})=\Omega_{n}(\boldsymbol{k})$. In strained TBG the Dirac cones are protected by $\mathcal{C}_{2z}$ and the Berry curvature is not well defined. To obtain a finite curvature, this symmetry must be broken. In TBG this can be achieved by suspension or encapsulation with hBN~\citep{Zhang2019b, CeaPantaGuinea2020}. The presence of hBN induces an staggered sublattice potential which results in a gap at the two Dirac cones with a finite Berry curvature. This gap is extremely sensitive to the degree of alignment between hBN and graphene and is nonzero even for a large misalignment~\citep{CeaPantaGuinea2020}. Theory and experiments suggest that the gap value varies between 0 and 30 meV~\citep{Hunt2013,San-Jose2014,Jung2015,Zhang2019b, CeaPantaGuinea2020}.

Following these considerations, we now examine the topological currents arising from the induced Berry curvature within the semiclassical Boltzmann transport theory. In the presence of an external electric field $\boldsymbol{\xi}$, in addition to the usual band dispersion contribution, an extra non-classical term also contributes to the velocity of the charge carriers~\citep{XiaoBerry2010}, $\hbar\boldsymbol{v}_{n}(\boldsymbol{k})=\nabla_{k}E_{n}(\boldsymbol{k})-e\boldsymbol{\xi}\times\boldsymbol{\Omega}_{n}(\boldsymbol{k})$,
where the second term is called anomalous velocity and is driven by a nonzero Berry curvature. This velocity is always transverse to the electric field and will give rise to a Hall current. Following Ref. \citep{TonyLowPaco2015}, the transverse currents up to second order in the electric field are $\mathcal{J^{\pm}}=\sigma_{0}^{\pm}\xi+\sigma_{1}^{\pm}\xi$ which can be separated by their propagation direction, $S\equiv\mathrm{sign}(\boldsymbol{v}_{n}\cdot\boldsymbol{\xi})$, where, $+1$  is for forward and $-1$ for backward propagating states. If we consider the $K_{+}$ valley, the conductivities are given by
\begin{align}
\sigma_{0}^{\pm} & =-\frac{2e^{2}}{\hbar}\sum_{n}\int_{S=\pm}d\boldsymbol{k}f_{0}(\boldsymbol{k})\Omega_{n}(\boldsymbol{k}),\label{eq: Conductbulk} \\
\sigma_{1}^{\pm} & =\frac{e^{3}\tau}{2\hbar^{2}}\sum_{n}\int_{S=\pm}d\boldsymbol{k}\left[\nabla_{\boldsymbol{k}}f_{0}(\boldsymbol{k})\cdot\boldsymbol{\xi}\right]\Omega_{n}(\boldsymbol{k}),
\label{eq: Conductsurf}
\end{align}
where $f_{0}$ is the Fermi-Dirac distribution which implicitly depends on the chemical potential $\mu$, $\tau$ is the scattering time and $n$ the band index. The linear contribution to the current, Eq.~\ref{eq: Conductbulk}, is a bulk phenomenon. This contribution is bounded by the Chern number if the integral is over an isolated band. The second term, Eq.~\ref{eq: Conductsurf}, is the non-linear contribution, where the gradient of the distribution function indicates that only the states close to the Fermi surface contribute to the integral. Fig.~\ref{fig: Figure1}e) illustrates the various topological transverse current components in a typical transport device geometry. 

\begin{figure*}
\begin{centering}
\includegraphics[scale=0.09]{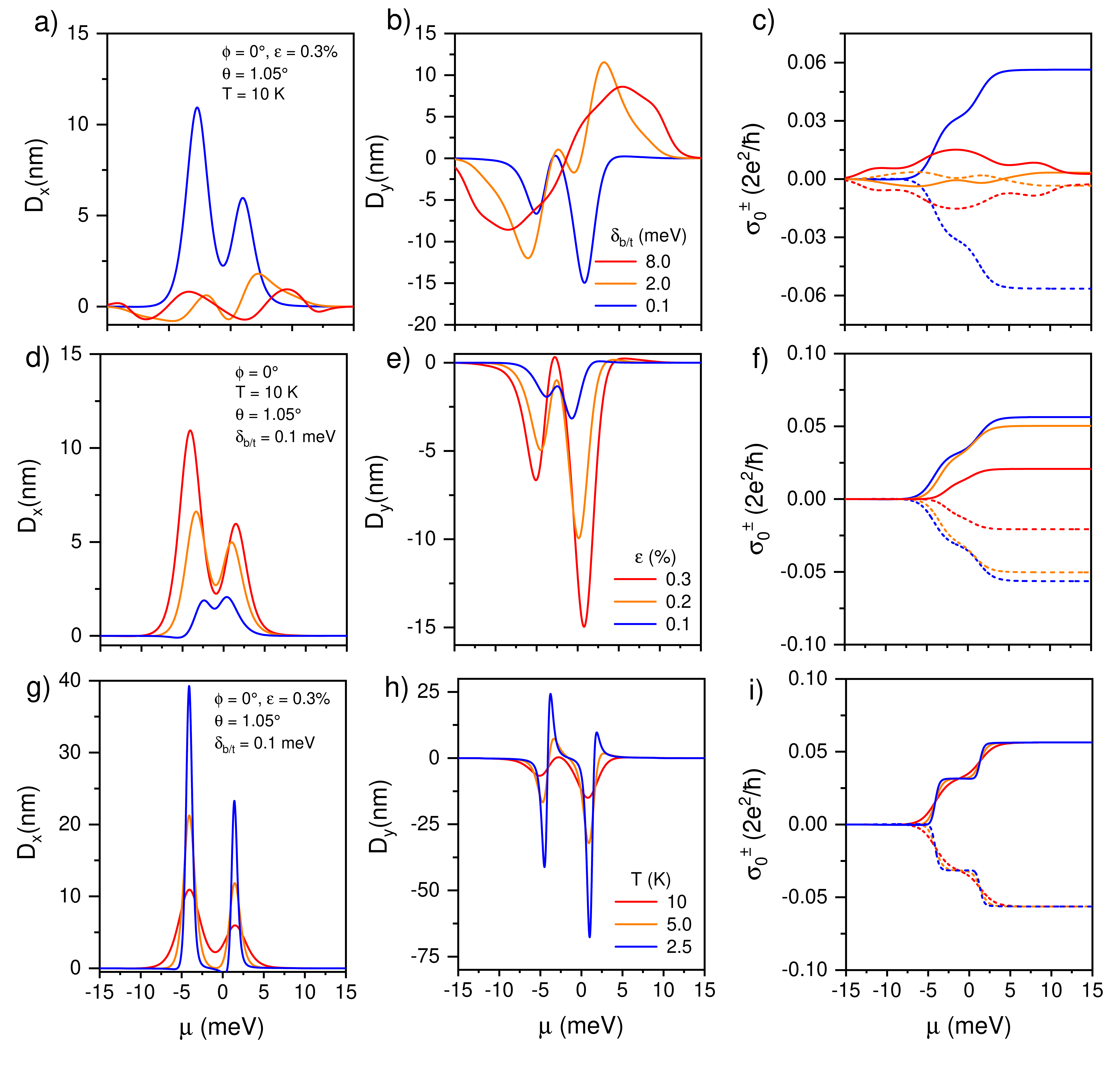}
\par\end{centering}
\caption{Dependence of the components of the Berry dipole and total Hall conductivity in strained twisted bilayer graphene on different external parameters: gap due to the substrate (top row), strain (central row), and temperature (bottom row). Panels in the first and second column are the components of the Berry dipole, panels in the third column are the forward (continuous line) and backward (dashed line) total Hall conductivity. The plots also show  the dependence on the chemical potential, whose range spans the two central bands. Hall conductivities in the third column are obtained by assuming an in-plane electric field parallel to the $x-$axis of the TBG unit cell.} 
\label{fig: Figure2}
\end{figure*}
The nonlinear Hall conductivity in a valley can be written as
\begin{equation}
\sigma_{1}^{\pm}=\frac{e^{3}\tau}{2\hbar^{2}}D^{\pm}\cdot\boldsymbol{\xi},
\label{eq: SurfaceConductivity}
\end{equation}
with $D^{\pm}=(D_{x}^{\pm},D_{y}^{\pm})$ the Berry dipole with
components
\begin{equation}
D_{\alpha}^{\pm}=\sum_{n}\int_{S=\pm}d\boldsymbol{k}\partial_{k_{\alpha}}f_{0}(\boldsymbol{k})\Omega_{n}(\boldsymbol{k}).\label{eq: Berry Dipole}
\end{equation}
The above equation measures the lowest-order correction to the total Berry curvature flux in the non-equilibrium state. If we consider the contribution from both valleys, time-reversal symmetry implies that both Berry curvature and anomalous velocity are odd in momentum. Notice that the group velocity is implicitly given in the derivative of the distribution function. Since the direction of the in-plane electric field is the same in each valley, we also have an inversion in the forward and backward propagating currents, $S\rightarrow-S$. Hence, the total and nonlinear Hall conductivities are respectively given by
\begin{align}
\sigma_{t,0}^{\pm} & =\sigma_{0}^{\pm}-\sigma_{0}^{\mp},\label{eq: TotalBulk}\\
\sigma_{t,1}^{\pm} & =\sigma_{1}^{\pm}+\sigma_{1}^{\mp}.\label{eq: TotalSurface}
\end{align}
In terms of the Berry dipole, the nonlinear Hall conductivity in Eq.~\ref{eq: TotalSurface}  can be written as 
\begin{equation}
\sigma_{t,1} =\frac{e^{3}\tau}{2\hbar^{2}}D\cdot\boldsymbol{\xi},
\label{eq: SurfDipoleTotal}
\end{equation}
where the total Berry dipole $D=D^{\pm}+D^{\mp}$ is independent of the electric field direction.  The nonlinear Hall current satisfy $\sigma_{t,1}^{+}=\sigma_{t,1}^{-}$ and is maximum when the Berry dipole is aligned with the electric field~\citep{SI}. 

\textit{Berry dipole and non-linear Hall effect.\textendash} Figure~[\ref{fig: Figure2a}] shows the Berry curvature of the valence and conduction, the two narrow bands in Fig.~[\ref{fig: Figure1}]d). The Berry curvature is highly concentrated near the anti-crossings of these bands. The finite temperature allows for the existence of a sizable Berry dipole when the chemical potential is not too close to these points. The Berry dipole as shown is two orders of magnitude larger than what has been reported in TMD and other van der Waals materials~\citep{Ma2019,KangMak2019,huang2020giant,hu2020nonlinear}. We attribute the origin of this large Berry dipole to the two almost touching quasi-flatbands, as illustrated in Fig. [\ref{fig: Figure1}]c) (see also Fig. [\ref{fig: Figure2a}]a)-b). As shown in Eq. \ref{eq: Berry Dipole}, the Berry dipole depends on the Berry curvature weighted by the derivative of the distribution function. In TMDs the bandwidth of the bands responsible of the Berry dipole usually have a large bandwidth and both Berry curvature and band velocities are spread in large regions of the mBZ, this results in a Berry dipole of a few Angstroms~\citep{YouLow2018}. In TBG, the narrow bands allow a concentration of both band velocity and Berry curvature at the vicinity of the gapped Dirac cones and near the magic angle where the bands are narrower and the Dirac cones more localized. As shown in Fig.~[\ref{fig: Figure2a}], the physical quantities defining Eq. \ref{eq: Berry Dipole} are quite concentrated in the same regions within the mBZ resulting in a large contribution to the Berry dipole.

\begin{figure*}
\begin{centering}
\includegraphics[scale=0.10]{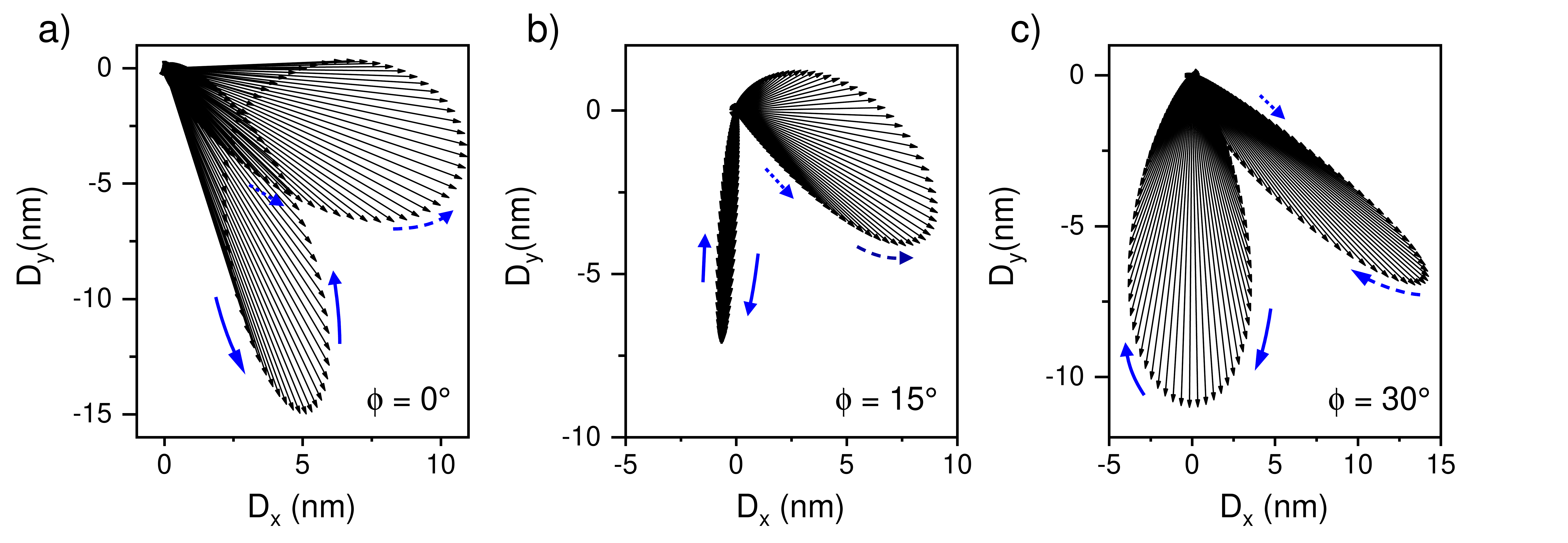}
\par\end{centering}
\caption{Berry dipole components of strained TBG as a function of the chemical potential and different strain directions. Black arrows indicate the dipole direction for each value of the chemical potential. Blue arrows represent the direction in which the chemical potential increases. Broken arrow shows the valence band, and full arrow shows the conduction band. Parameters: $\epsilon=0.3\%$, $\delta_{b/t}=0.1$ meV, $\theta=1.05^\circ$ and $T=10$ K}
 \label{fig: Figure3}
\end{figure*}

Figure [\ref{fig: Figure2}] surveys the tunability of the Berry dipole as function of the strain parameters, electron fillings, gap size, and temperature. The calculated Berry dipole is of order $D\sim 10\,$nm or larger over the range of parameters used in these calculations. We notice that the linear contribution in Eq. \ref{eq: Conductbulk} is bounded by the total Berry curvature of the narrow bands while the non-linear contribution in Eq. \ref{eq: Conductsurf} can be modified by both strain and mass gap. As shown in Fig. [\ref{fig: Figure2}] the Berry dipole is enhanced by reducing the mass gap, temperature or by modifying the strain magnitude or direction (see also Fig. [\ref{fig: Figure3}]). This allow us to obtain an estimate of the crossover electric field, ${\cal E}_c$, for which the non linear $\vec{j}_{NL} \sim \vec{\cal E}^2$, transverse current becomes comparable to the linear parallel current, $\vec{j}_L \sim \vec{\cal E}$. We obtain ${\cal E}_c \sim ( \hbar v_F k_F) / ( e D )$, where $D$ is the Berry dipole, $v_F$ is an average of the band velocity at the Fermi level, and $k_F$ is an average of the Fermi wavevector. For $\hbar v_F k_F \sim 1 - 10$ meV, we obtain ${\cal E}_c \sim 10^{-4} - 10^{-5}$ mV nm$^{-1}$. Hence, it is most likely that in such samples (for example TBG near the magic angle with an unaligned hBN substrate as in Ref.~\cite{Jetal19,Choi2019,Xie2019,Kerelski2019,Quiao2018ht,Huder}), a charge Hall response will be observed whose Hall voltage depends on the longitudinal electric field as ${\cal E}^2$. Note that, if the system is driven at a finite frequency, the non linear response will lead to a significant dc current.

Figure [\ref{fig: Figure3}] plots the Berry dipole vectors for varying chemical potential across the two quasi-flatbands, for uniaxial strains along different crystallographic directions as indicated. The two “fans” correspond to the two quasi-flatbands with zero dipole magnitude when the TBG is at half-filling. The Berry dipole vectors rotate in a clockwise or anti-clockwise manner with increasing chemical potential for uniaxial strains applied along the different directions. Furthermore, since the Berry dipole is a Fermi surface effect, the effect is strongest when the chemical potential leads to half filling of either of the two quasi-flatbands, i.e. 1/4 or 3/4 filling. In transport experiments, maximal anomalous Hall effect corresponds to the situation when the in-plane electric field is aligned with the Berry dipole vector. Hence, experimental design of the transport experiment should take into account both the uniaxial strain direction and the chemical potential. These calculations suggest that the chemical potential can allow for a tunability of the Berry dipole vector by about $45^o$.

{\it Conclusions.}
We theoretically study the emergence of topological currents in twisted bilayer graphene with uniaxial strain. The non-linear Hall contribution reported here,  Eq.~\ref{eq: SurfDipoleTotal} arises even in the presence of time-reversal but requires an in-plane electric field~\citep{SodemannLiang2015,TonyLowPaco2015,YuAny2014}. This non-linear Hall  response can be significantly larger than in other two dimensional materials, due to the enhanced Berry dipole of order $10\,$nm. Its value is highly sensitive to the amount of heterostrain, and to the value of the gap opened by the substrate. Parameters such as the width of the central bands, the value of the strains, and the band gap, induced by the alignment with the hBN substrate, can be tuned experimentally~\cite{disorder}. 

The linear term in Eq. \ref{eq: TotalBulk} leads to a charge neutral current which can be detected by non-local transport measurements~\citep{Getal14,SuiValley2015,Wu2019NonLocal} or by explicitly breaking time-reversal symmetry~\citep{Klitzing1980,NagaosaHall2010}. It has been proposed that the high resistivity states phases observed near integer fillings break the equivalence between valleys~\cite{XM20,CG20,BKLCVZ20,Wong2019b,Zondiner2019b}.
Then, the currents calculated in Eq.~\ref{eq: TotalBulk} give an estimate of a finite Hall conductivity~\cite{Serlin2019} in the absence of a magnetic field. Hence, dc transport measurements can be used to infer the nature of correlated phases.

Finally, it is worth noting that both strains and the lack of inversion symmetry induced by the substrate favor an increase in the Berry curvature, leading to the effects discussed here. Eventually, these perturbations distort and broaden the central bands of TBG. We consider that the values for the strain and substrate induced gap used in this work give a reasonable compromise where the topological features discussed here can be observed.

{\it As this manuscript was approaching completion, a manuscript\cite{ZXZX20} has been posted, arXiv:2010.08333, addressing similar topics. As far as the manuscripts overlap, the results are qualitatively consistent. The values of the Berry dipole reported are somewhat different. This difference can be ascribed to the significant dependence of this value on parameters external to twisted bilayer graphene, the magnitude of the strain, and, particularly, the sublattice asymmetry induced by the substrate, as discussed here.}

{\it Acknowledgements.} P.A.P. and F. G. acknowledge funding from the European Commision, under the Graphene Flagship, Core 3, grant number 881603, and from grants NMAT2D (Comunidad de Madrid, Spain),  SprQuMat and SEV-2016-0686, (Ministerio de Ciencia e Innovación, Spain). T.L. acknowledges support from the National Science Foundation under Grant No. NSF/EFRI-1741660.

\balance
\bibliographystyle{apsrev4-1}

\clearpage

\onecolumngrid
\begin{center}
\Large Supplementary information for \\ Tunable large Berry dipole in strained twisted bilayer graphene
\end{center}
\begin{center}
Pierre A. Pantale\'on, Tony Low and Francisco Guinea
\end{center}

\setcounter{equation}{0}
\setcounter{figure}{0}
\setcounter{table}{0}
\setcounter{page}{1}
\makeatletter
\renewcommand{\theequation}{S\arabic{equation}}
\renewcommand{\thefigure}{S\arabic{figure}}

\section{Twisted Bilayer Graphene with uniaxial strain}

In a monolayer graphene the primitive lattice vectors are $a_{1}=a(1,0)$ and $a_{2}=a(1/2,\sqrt{3}/2)$ with $a\approx2.46 \text{\AA}$ the lattice constant. The reciprocal lattice vectors, $b_{i}$, $i=1,2$ satisfying $a_{i}\cdot b_{j}=2\pi\delta_{ij}$ are then given by $b_{1}=\frac{2\pi}{a}(1,-1/\sqrt{3})$ and $b_{2}=\frac{2\pi}{a}(1,-1/\sqrt{3})$. The graphene Dirac cones are located at $\boldsymbol{K}_{\pm}=-\xi(2b_{1}+b_{2})/3$ with $\xi=\pm1$ a valley index. For a twisted bilayer graphene, we define the structure as in Ref. \citep{Moon2013}, by rotating layers $l=1$ and $l=2$ of the $AA$-stacked configuration around a common $B$ site by $-\theta/2$ and $\theta/2$, respectively. If $R(\theta)$ represents a rotation matrix by $\theta$, the primitive and reciprocal lattice vectors in each rotated layer are written as $a_{i}^{(l)}=R(\mp\theta/2)a_{i}$ and $b_{i}^{(l)}=R(\mp\theta/2)b_{i}$, respectively.  We now introduce a geometric uniaxial deformation, where the bilayer system is relatively stressed along one direction and unstressed on the perpendicular direction~\citep{BiFu2019}. Geometrically, uniaxial strain can be described by two parameters, the strain relative magnitude $\epsilon$ and the strain direction $\phi$. The strain tensor $\mathfrak{\varepsilon}$ in terms of these two parameters is written as 

\begin{equation}
\varepsilon=\epsilon\left(\begin{array}{cc}
-\cos^{2}\phi+\nu\sin^{2}\phi & (1+\nu)\cos\phi\sin\phi\\
(1+\nu)\cos\phi\sin\phi & -\sin^{2}\phi+\nu\cos^{2}\phi
\end{array}\right),\label{eq: StrainTensor}
\end{equation}
where $\nu=0.16$ is the Poisson ratio for graphene. In strained TBG, the transformed primitive and reciprocal lattice vectors for each rotated graphene layer are given by 

\begin{align}
\alpha_{i}^{(l)} & =(\mathbb{I}+\varepsilon_{l})a_{i}^{(l)},\label{eq: StrainRealPrimitive}\\
\beta_{i}^{(l)} & =(\mathbb{I}-\varepsilon_{l}^{T})b_{i}^{(l)}.\nonumber 
\end{align}
with $\varepsilon_{l}$ the strain tensor and $l$ a layer index. In TBG with uniaxial heterostrain, the relative deformation satisfy $\varepsilon=\varepsilon_{2}-\varepsilon_{1}$ with $\varepsilon_{2} = -\varepsilon_{1} =\frac{1}{2}\varepsilon$. This ensure that each layer is strained oppositely with the same magnitude \cite{BiFu2019}. As shown in Fig.~[1a)], in TBG without strain, the moir\'e dots in the lattice have a circular shape. The combined effect of twist and strain deforms the lattice and these dots become elliptical, Fig.~[1b)]. In this situation and by Eq.~\ref{eq: StrainRealPrimitive}, the reciprocal lattice vectors of the deformed moir\'e superlattice are $\boldsymbol{g}_{i}=\beta_{i}^{(1)}-\beta_{i}^{(2)}$.

In addition to the geometrical effects, the strain affects the electronic structure by a change in the intralayer Hamiltonian. In the small strain limit, this change can be introduced by a pseudomagnetic field in the low energy Hamiltonian~\citep{Nam2017}. In each graphene layer the vector potential, $A^{(l)}=(A_{x}^{(l)},A_{y}^{(l)})$, for this field is given by \citep{GuineaStrain} 

\begin{align}
A_{x}^{(l)} & =\frac{\sqrt{3}}{2a}\beta_{G}[\varepsilon_{xx}^{(l)}-\varepsilon_{yy}^{(l)}],\nonumber \\
A_{y}^{(l)} & =\frac{\sqrt{3}}{2a}\beta_{G}[-2\varepsilon_{xy}^{(l)}],\label{eq:VectorPotGrunn}
\end{align}
where $\beta_{G}\approx3.14$ is a dimensionless parameter and $l$ a layer index. In TBG with small twist, the moir\'e superlattice constant is much larger than the atomic scale and the low energy is dominated by states near points $K_{+}$ and $K_{-}$, therefore we can analyze
each valley separately. The low energy Hamiltonian for TBG with uniaxial strain can be written as

\begin{equation}
H=\left(\begin{array}{cc}
H(\boldsymbol{q}_{1,\zeta})+\delta_{b}\sigma_{z} & U^{\dagger}\\
U & H(\boldsymbol{q}_{2,\zeta})+\delta_{t}\sigma_{z}
\end{array}\right),\label{eq: HamiltonianStrainedTBG}
\end{equation}
where $\boldsymbol{q}_{l,\zeta}=R(\pm\theta/2)(\mathbb{I}+\varepsilon_{l}^{T})(\boldsymbol{q}-D_{l,\zeta})$
with $\pm$ for $l=1$ and $l=2$, respectively. $D_{l,\zeta}=(\mathbb{I}-\varepsilon^{T})\boldsymbol{K}_{\xi}-\zeta A^{(l)}$ are the re-scaled valley points and $H(\boldsymbol{q})=-(\hbar v_{f}/a)\boldsymbol{q}\cdot(\zeta\sigma_{x},\sigma_{y})$ is the Hamiltonian for a monolayer graphene, $\zeta=\pm1$ is a valley index. In the above equation, $U$ is the interlayer coupling between twisted graphene layers given by the Fourier expansion, 

\begin{align}
U & =\left(\begin{array}{cc}
u & u^{\prime}\\
u^{\prime} & u
\end{array}\right)+\left(\begin{array}{cc}
u & u^{\prime}\omega^{-\zeta}\\
u^{\prime}\omega^{\zeta} & u
\end{array}\right)e^{i\zeta\boldsymbol{g}_{1}\cdot\boldsymbol{r}}\nonumber 
  +\left(\begin{array}{cc}
u & u^{\prime}\omega^{\zeta}\\
u^{\prime}\omega^{-\zeta} & u
\end{array}\right)e^{i\zeta(\boldsymbol{g}_{1}+\boldsymbol{g}_{2})\cdot\boldsymbol{r}}
\label{eq: InterlayerU}
\end{align}
where $\omega=e^{2\pi i/3}$, with $u=0.0797$ eV and $u^{\prime}=0.0975$~eV~\citep{Koshino2018a} the amplitudes which take into account out-of-plane corrugation effects~\citep{Koshino2018a,TKV19,Nam2017}. Lastly, $\delta_{b/t}$ is an staggered sublattice potential which takes into account the effects of encapsulation with an hBN substrate~\citep{CeaPantaGuinea2020}. To obtain the energy spectrum, the secular equation to be solved is $H(\boldsymbol{k})\left|\Psi_{n\boldsymbol{k}}\right\rangle =E_{n}(\boldsymbol{k})\left|\Psi_{n\boldsymbol{k}}\right\rangle $
where $\left|\Psi_{n\boldsymbol{k}}\right\rangle $ and $E_{n}(\boldsymbol{k})$
are the eigenvectors and eigenenergies. The momentum in Eq.~\ref{eq: HamiltonianStrainedTBG}
is $\boldsymbol{q}=\boldsymbol{k}+n\boldsymbol{g}_{1}+m\boldsymbol{g}_{2}$,
with $m,n$ integers, and $\boldsymbol{k}$ is restricted to the first mBZ. In the numerical calculations, the number of Fourier components defining the eigenvectors is bounded by a cutoff: $|\boldsymbol{g}|<\boldsymbol{g}_{c}$, where $\boldsymbol{g}_{c}$ is chosen in order to achieve the convergence of the low energy bands. 

\section{Additional Results}
\begin{figure*}
\begin{centering}
\includegraphics[scale=0.11]{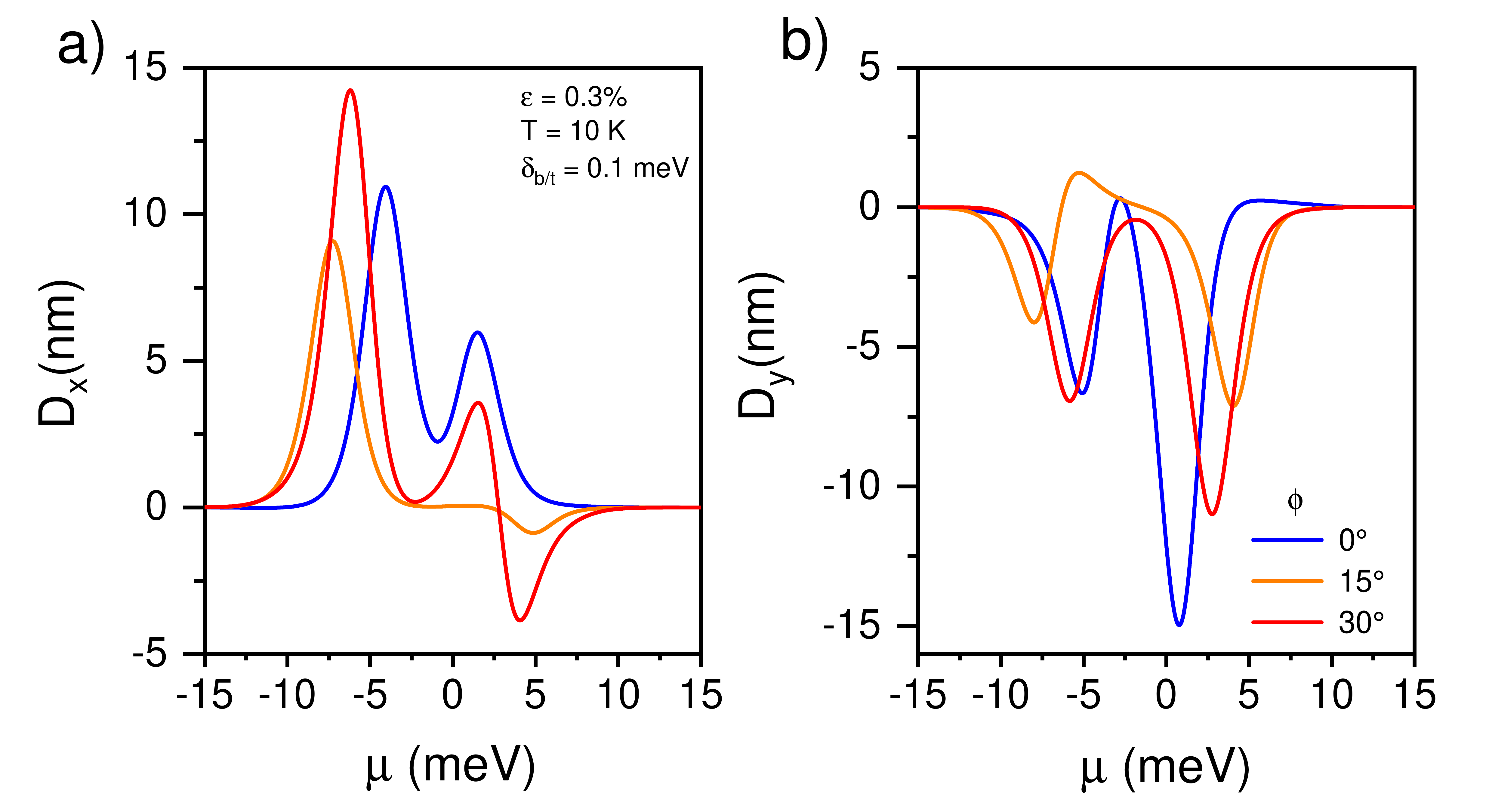}
\par\end{centering}
\caption{Berry dipole components of TBG as a function of the chemical potential. a) $D_{x}$ and b) $D_{y}$ are the Berry dipole components calculated for different strain directions. 
\label{fig: FigureS1}}
\end{figure*}

\begin{figure*}
\begin{centering}
\includegraphics[scale=0.11]{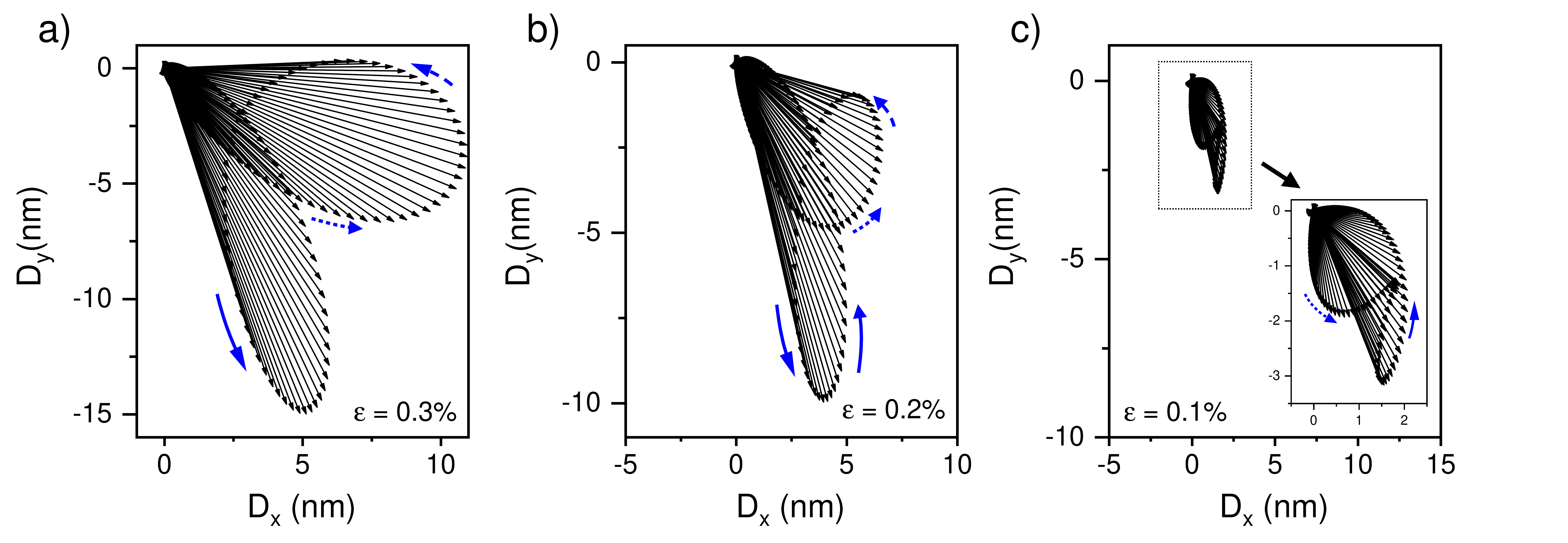}
\par\end{centering}
\caption{Berry dipole components of strained TBG  as a function of the chemical potential and different strain directions. Black arrows indicate the dipole direction for each value of the chemical potential. Blue arrows represent the direction in which the chemical potential increases. Broken arrow shows the valence band, and full arrow shows the conduction band.  Parameters are the same as in Fig. [3d)-e)]}
\label{fig: FigureS2}
\end{figure*}

\begin{figure*}
\begin{centering}
\includegraphics[scale=0.11]{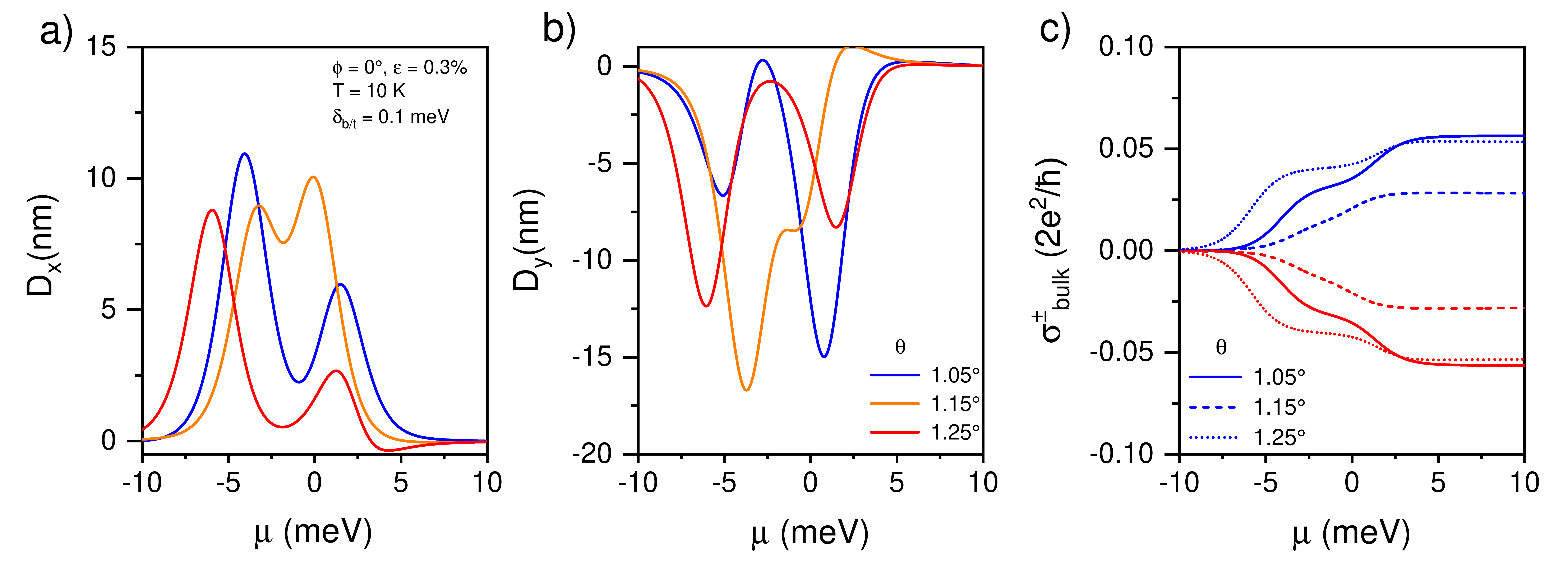}
\par\end{centering}
\caption{Berry dipole components of TBG as a function of the chemical potential and different twist angles. In panels a) and b) we display the Berry dipole components and c) is the total bulk Hall conductivity. Continuous (dotted) lines represent forward (backward propagating states). In this case, the in-plane electric field is parallel to the x-axis. 
\label{fig: FigureS3}}
\end{figure*}

Figure~[\ref{fig: FigureS1}] shows the Berry dipole components as a function of the chemical potential for different strain directions. Is clear that the magnitude and direction of the Berry dipole is modified by the direction of the applied strain. Figure \ref{fig: FigureS2} shows the Berry dipole for different strain values. Black arrows indicate the dipole direction for each value of the chemical potential. Blue arrows indicate the direction in which the chemical potential increases.  Broken arrow shows the valence band, and full arrow shows the conduction band. Here, as the strain magnitude decreases, the magnitude of the Berry dipole also decreases. Figure~[\ref{fig: FigureS3}] shows the Berry dipole components as a function of the chemical potential for different twist angles. 

\section{Topological phases induced by the substrate}
\begin{figure*}
\begin{centering}
\includegraphics[scale=0.11]{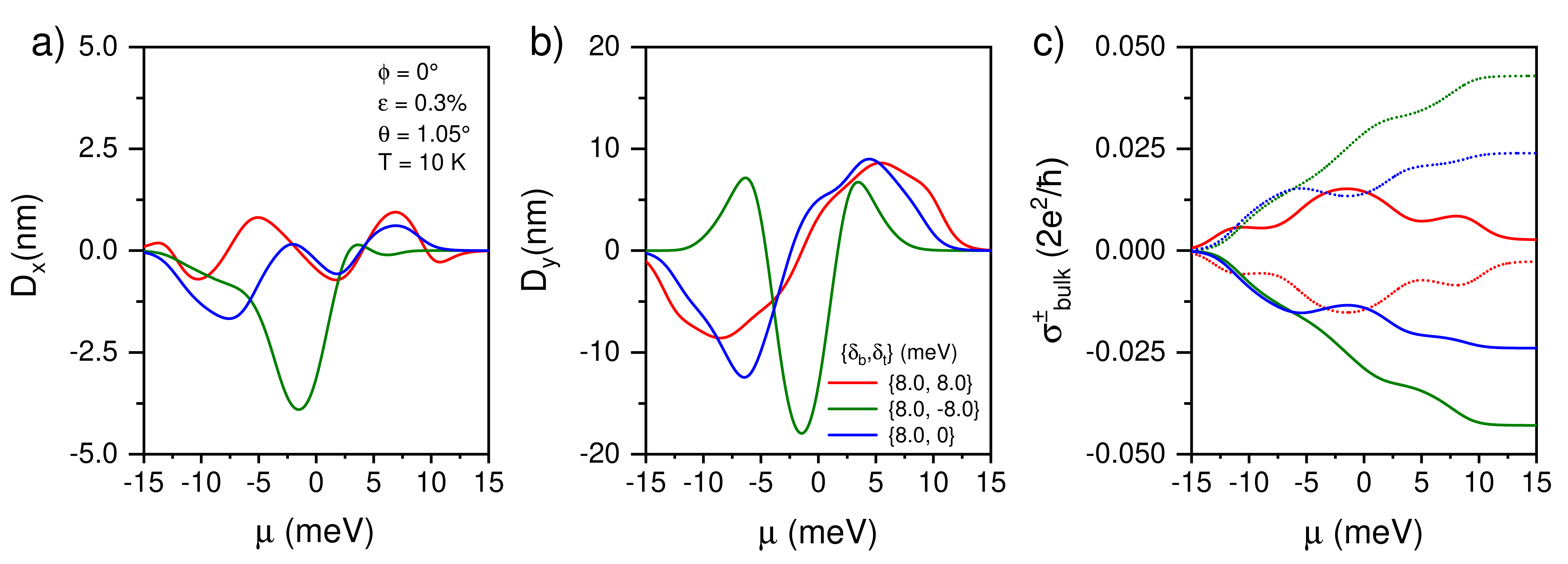}
\par\end{centering}
\caption{Berry dipole components of TBG as a function of the chemical potential. By suspending or encapsulating TBG with hBN, the values of the staggered potential can be modified. Panels a) and b) display the Berry dipole components and c) is the total bulk Hall conductivity. The values of the staggered potential are shown in b). Continuous (dotted) lines represent forward (backward propagating states). In this case, the in-plane electric field is parallel to the x-axis.    
\label{fig: FigureS4}}
\end{figure*}

By suspending or encapsulating TBG with hexagonal boron nitride different topological phases can be obtained~\citep{CeaPantaGuinea2020}. Our calculations found that for small values of the sublattice potential the system has two topological phases. The phase with Chern numbers $\mathcal{C}=\pm1$, is obtained in the cases of $\delta_{b}\neq0, \delta_{t}=0$ (suspended TBG/hBN) and $\delta_{b}=\delta_{t}$ (encapsulated hBN/TBG/hBN). The additional phase with Chern number $\mathcal{C}=0$ in both narrow bands is obtained for $\delta_{b}=-\delta_{t}$ (encapsulated hBN/TBG/hBN). Figure~\ref{fig: FigureS4} shows the Berry dipole components and the total Hall conductivity for different combinations of the sublattice potential. In the phase with non-zero Chern numbers the behavior of the Berry dipole components as a function of the chemical potential is quite similar (red and blue lines) in both cases. In the phase with zero Chern number the sign of the Berry curvature is opposite at the Dirac cones and the magnitude of the Berry dipole components is modified. However, in both phases, the magnitude of the Berry dipole is of the same order. 

\section{Numerical evaluation of the Hall conductivity}

In this section, we address the numerical evaluation of the Hall conductivity where the breaking of inversion symmetry in the single valley model in Eq.~\ref{eq: HamiltonianStrainedTBG} allows for a finite Berry curvature
\begin{equation}
\overrightarrow{\Omega}_{n}(\boldsymbol{k})=\nabla_{\boldsymbol{k}}\times\boldsymbol{A}_{n}(\boldsymbol{k}),\:\boldsymbol{A}_{n}(\boldsymbol{k})=i\left\langle \Psi_{nk}\right|\nabla_{\boldsymbol{k}}|\left. \Psi_{nk}\right\rangle ,
\label{eq:Omega}
\end{equation}
 where $n$ is the band index, $\boldsymbol{A}_{n}(\boldsymbol{k})$
is the Berry connection and $\Psi_{nk}$ the eigenvectors of Eq.~\ref{eq: HamiltonianStrainedTBG}. In the numerical evaluation of Eq.~3 and Eq.~4 we first define a Monkhorst-Pack grid in the mBZ in Fig. 1c). In our calculations we use a grid size of $3N^{2}$ $\boldsymbol{k}$-points, with $N\sim600-700$. Next we calculate $\Omega_{n}(\boldsymbol{k})$ following a procedure similar to that in Ref.~\citep{Fukui2005}, where we numerically integrate the Berry connection  $\boldsymbol{A}_{n}(\boldsymbol{k})$ in small loops around each momentum $\boldsymbol{k}$. For each loop we choose a set of eigenvectors around the loop and then we calculate the Berry connection between points of the loop. The total contribution of each small loop is the local Berry curvature determined up to a factor of $2\pi$. Rescaling by $2\pi$ gives the Chern number contribution from each loop, the sum of which over the Brillouin zone gives the Chern number of the corresponding band. The integrals in Eq.~3 and Eq.~4 are obtained by summing the Berry curvature at each point weighted by the corresponding scalar function. This numerical procedure allow us to eliminate numerical problems with gauge choices because the arbitrary phases appear twice with opposite sign. However, depending on the distribution of the Berry curvature within the mBZ, the size of the loop must be small enough to achieve accurate numerical results. 

\end{document}